\documentclass[aps,pra,twocolumn,showpacs,superscriptaddress,longbibliography]{revtex4-1}
\usepackage{enumerate}
\usepackage{graphicx}
\usepackage{amsmath}
\usepackage{mathtools}
\usepackage{amsfonts}
\usepackage{amssymb}
\newcommand{\ket}[1]{\left| #1 \right\rangle}
\newcommand{\bra}[1]{\left\langle #1 \right|}

\begin{document}
\today
\title{Scalable star-shape architecture for universal spin-based nonadiabatic holonomic quantum computation}
\author{Vahid Azimi Mousolou\footnote{Electronic address: v.azimi@sci.ui.ac.ir}}
\affiliation{Department of Mathematics, Faculty of Science, University of Isfahan, Box 81745-163 Isfahan, Iran}
\affiliation{School of Mathematics, Institute for Research in Fundamental Sciences (IPM), P. O. Box 19395-5746, Tehran, Iran}
\begin{abstract}
 Nonadiabatic holonomic quantum computation as one of the key steps to achieve fault tolerant quantum information processing has so far been realized in a number of physical settings. However, in some physical systems particularly in spin qubit systems, which are actively considered for realization of quantum computers, experimental challenges are undeniable and the lack of a practically feasible and scalable scheme that  supports universal holonomic quantum computation all in a single well defined setup is still an issue.
Here, we propose and discuss a scalable star-shape architecture with promising feasibility, which may open up for realization of universal (electron-)spin-based nonadiabatic holonomic quantum computation. 
  
\end{abstract}
\pacs{}
\maketitle

\section{Introduction}

Holonomic quantum computation \cite{zanardi99,xiang-bin2001,zhu2002,sjoqvist2012} is recognised among key
approaches to fault resistant quantum computation. Nonadiabatic holonomic quantum computation \cite{xiang-bin2001,zhu2002,sjoqvist2012} 
compared to its adiabatic counterpart \cite{zanardi99} is more compatible with the short coherence time of quantum bits (qubits). 
To achieve a feasible platform, nonadiabatic holonomic quantum computation has been adapted and developed for different physical settings
\cite{xiang-bin2001,zhu2002, zhu2003a, zhu2003b, sjoqvist2012, mousolou2014, mousolou2017a, mousolou2017b, zhao2017b}. Nonadiabatic 
holonomic quantum computation has also been combined with decoherence free subspaces 
\cite{xu2012, xu2014a, liang2014, xue2015, zhou2015, xue2016, zhao2017a, mousolou2018}, 
noiseless subsystems \cite{zhang2014}, and dynamical decoupling \cite{xu2014b} to further improve its robustness.
Experimental realizations of nonadiabatic holonomic quantum computation in various physical systems, such 
as NMR \cite{feng2013, li2017}, superconducting transmon \cite{Abdumalikov2013, danilin2018, eggers2018}, and 
NV centers in diamond \cite{Arroyo-Camejo2014, Zu2014, Zhou2017, sekiguchi2017, ishida2018} have been carried out. 

Nevertheless, the implementation of nonadiabatic holonomic quantum computation in some physical systems particularly in spin qubit system, 
which is one of the natural and promising candidates to built quantum computers upon, has been remained at the level of single-qubit gates.       
In fact, from practical perspectives, establishing a scalable multipartite scheme, which possesses full holonomic computational power for quantum 
processing, in these physical systems is still a challenge.

In this paper, we aim to address this issue by proposing a scalable architecture for universal spin-based nonadiabatic holonomic quantum computation, 
which enjoys a reasonable capability of being implemented with current technologies. We consider a star-shape system, where in principal an arbitrary number of register spin qubits are arranged about and all coupled to an auxiliary spin qubit in the middle of architecture. Universal holonomic single-qubit gates are achieved by controlling the coupling between two computational basis states of register qubits through local transverse magnetic fields. The middle auxiliary spin qubit introduces an indirect bridge coupling between each pair of register qubits bringing about a double $\Lambda$ structure, which permits to implement holonomic entangling gates between selected pair.  

\section{Scalable architecture}

\begin{figure}[h]
\begin{center}
\includegraphics[width=7.5 cm]{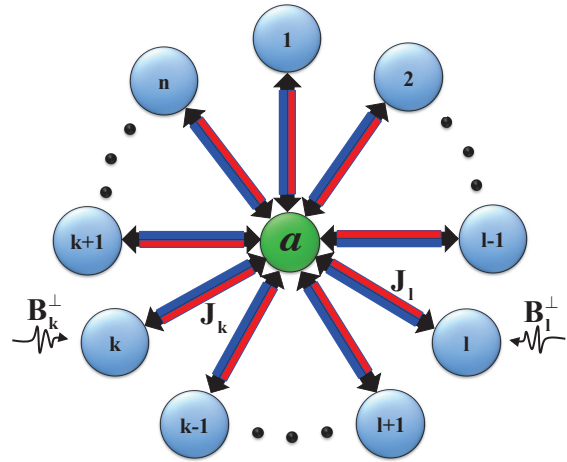}
\end{center}
\caption{(Color online) Scalable star-shape architecture for universal spin-based nonadiabatic holonomic quantum computation. An arbitrary $n$ number of register spin qubits  arranged about and all coupled to an auxiliary spin qubit deployed in the middle of architecture. Each register qubit is allowed to interact with a controllable local magnetic field.}
\label{fig:star-shape}
\end{figure}

The model system that we have in mind is a scalable $n$ register spin qubits coupled in a star-shape architecture through an auxiliary  spin qubit as depicted in Fig. \ref{fig:star-shape}.

Since only universal single-qubit and two-qubit gates are needed to achieve a universal quantum information processing, the Hamiltonian adapted here is a collective single-qubit and two-qubit Hamiltonians given by 
\begin{eqnarray}
H=\sum_{k=1}^{n}H_{k}+\sum_{k, l=1}^{n}H_{kl}.
\label{CH}
\end{eqnarray}
For single-qubit Hamiltonians we consider 
\begin{eqnarray}
H_{k}=B_{k}^{\bot}\cdot S^{(k)},
\end{eqnarray}
which describes the interaction of the $k$th spin qubit, $S^{(k)}=(S^{(k)}_{x} ,S^{(k)}_{y} ,S^{(k)}_{z})$, with a local transverse magnetic field $B_{k}^{\bot}=(B_{k}^{x}, B_{k}^{y}, 0)$.

Two-qubit Hamiltonians read
\begin{eqnarray}
H_{kl}=J_{k} H^{(k)}_{XY}+J_{l} H^{(l)}_{XY},
\label{tqh}
\end{eqnarray}
where $H^{(\bullet)}_{XY}=S^{(\bullet)}_{x}S^{(a)}_{x}+S^{(\bullet)}_{y}S^{(a)}_{y}$, bullet stands for the corresponding superscript $k$ or $l$, and $a$ represents the auxiliary qubit. The $H_{kl}$ describes a three-body anisotropic interaction between the corresponding register qubits $k, l$, and the auxiliary qubit. The $J_{k}$ and $J_{l}$ are the exchange coupling strengths to the auxiliary qubit. In fact, the two-qubit Hamiltonian in Eq. (\ref{tqh}) introduces an indirect coupling between the selected two register qubits $k$ and $l$ through the auxiliary qubit. This can be seen in Fig. \ref{fig:star-shape} as well.

In the following, we discuss the realization of a universal family of single-qubit and two-qubit gates in this setup.

\subsection{Singel-Qubit Gates}

For a single-qubit gate on the given qubit $k$, we only turn on the single-qubit Hamiltonian, $H_{k}$, in the collective Hamiltonian given in Eq. (\ref{CH}) by 
exposing the qubit $k$ to a local transverse magnetic field $B_{k}^{\bot}$. During this implementation, we assume that the other terms in Eq. (\ref{CH}) are kept off.
 Thus, our effective Hamiltonian in this case is 
 \begin{eqnarray}
H_{k}=B_{k}^{\bot}\cdot S^{(k)}=\frac{B}{2}\vec{n}\cdot\vec{\sigma},
\end{eqnarray}
where $B$ and $\vec{n}=(\cos\beta, \sin\beta, 0)$, respectively, describe the strength and the direction of the local transverse magnetic field $B_{k}^{\bot}$ in the $xy$ plane. 
The $\vec{\sigma}=(\sigma_{x}, \sigma_{y}, \sigma_{z})$ is the standard Pauli operators and $\hbar=1$ from now on.

To achieve holonomic single-qubit gates, we consider cyclic evolutions of an arbitrary qubit state 
 \begin{eqnarray}
\ket{\psi}=\cos\frac{\theta}{2}\ket{0}+e^{i\phi}\sin\frac{\theta}{2}\ket{1},
\end{eqnarray}
in which only geometric phases are relevant. Explicitly, we are interested in evolutions 
 \begin{eqnarray}
 \mathcal{U}(\tau_{0}, \tau)\ket{\psi}=\exp[-i\int_{\tau_{0}}^{\tau}H_{k}(s)ds]\ket{\psi}
 \end{eqnarray}
, along which no dynamical phases occur, for instance the condition
   \begin{eqnarray}
 \bra{\psi}\mathcal{U}^{\dagger}(\tau_{0}, t)H_{k}(t) \mathcal{U}(\tau_{0}, t)\ket{\psi}=0
 \label{DC}
 \end{eqnarray}
is satisfied at any time $t\in[\tau_{0}, \tau]$ \cite{aharonov1987}. Considering a local transverse magnetic field $B_{k}^{\bot}=B\vec{n}$ with constant phase $\beta$, reduces the condition in Eq. (\ref{DC}) to one of the following simplified conditions:
 \begin{eqnarray}
(i)\ && \phi-\beta=(2m+1)\frac{\pi}{2},\ \ \ \ \ \ \  m=0, \pm 1, \pm2, ... \nonumber\\ 
(ii) \ &&\theta=0\  \text{or}\ \pi
\label{SDC}
\end{eqnarray}
This follows from the fact that $[H_{k}(t), \mathcal{U}(\tau_{0}, t)]=0$ at any time $t$, when the phase $\beta$ is fixed constant.

In the light of the above simplified conditions, we carry out our cyclic evolution in the following three steps:
\begin{itemize}
\item Step 1: We first evolve the general initial state $\ket{\psi}$ to the computational basis state $\ket{0}$ by turning on the local transverse magnetic field $B_{k}^{\bot}=B\vec{n}$ for a time interval $[0, \tau_{1}]$ with constant phase $\beta=\phi-\frac{\pi}{2}$ and (time-dependent) strength  $B$ such that $\int_{0}^{\tau_{1}}Bdt=\theta$. Hence, we have 
 \begin{eqnarray}
\mathcal{U}(0, \tau_{1})\ket{\psi}=\ket{0}.
 \end{eqnarray}
 \item Step 2: Next, we evolve the state $\ket{0}$, all the way along the meridian of the Bloch sphere corresponding to the fixed azimuthal angle $\tilde{\phi}$, to the state $e^{i\tilde{\phi}}\ket{1}$ by employing the constant magnetic phase $\beta=\tilde{\phi}+\pi/2$ and (time-dependent) magnetic strength  $B$ for a time interval $[\tau_{1}, \tau_{2}]$ such that $\int_{\tau_{1}}^{\tau_{2}}Bdt=\pi$. Namely,
    \begin{eqnarray}
\mathcal{U}(\tau_{1}, \tau_{2})\ket{0}=e^{i\tilde{\phi}}\ket{1}.
 \end{eqnarray}
 \item Step 3: Finally, we run the Hamiltonian $H_{k}$ for another time interval $[\tau_{2}, \tau_{3}]$ with fixed magnetic phase $\beta=\phi-\pi/2$
 and (time-dependent) magnetic strength  $B$ such that $\int_{\tau_{2}}^{\tau_{3}}Bdt=\pi-\theta$. This would evolve the final state of step 2, i.e., the state $e^{i\tilde{\phi}}\ket{1}$, into the final state 
 $e^{i\Delta\phi}\ket{\psi}$, where $\Delta\phi=\tilde{\phi}-\phi$. In other words
\begin{eqnarray}
\\\mathcal{U}(\tau_{2}, \tau_{3})e^{i\tilde{\phi}}\ket{1}=e^{i\Delta\phi}\ket{\psi}.
\end{eqnarray} 
\end{itemize}

 \begin{figure}[h]
\begin{center}
\includegraphics[width=6cm]{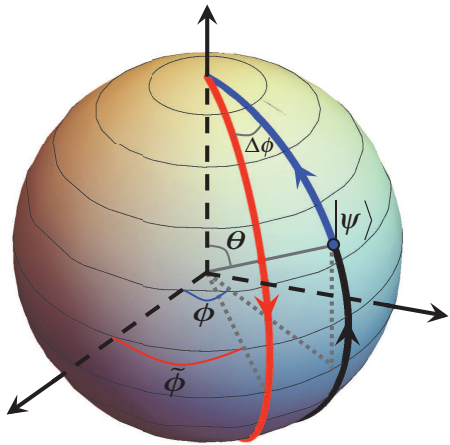}
\end{center}
\caption{(Color online) Cyclic evolution of a general qubit state $\ket{\psi}$ on Bloch sphere is carried out in three steps: step 1, which is illustrated in blue, evolves the state $\ket{\psi}$ to the state $\ket{0}$ along the meridian of the Bloch sphere corresponding to the fixed azimuthal angle $\phi$; step 2, which is shown in red, moves the north pole state $\ket{0}$ all the way down to the south pole state $e^{i\tilde{\phi}}\ket{1}$ along the meridian of the Bloch sphere corresponding to the fixed azimuthal angle $\tilde{\phi}$; finally, step 3, which is depicted in black, evolves the state $e^{i\tilde{\phi}}\ket{1}$ back into the initial state with overall accumulated phase $\Delta\phi=\tilde{\phi}-\phi$, i.e., $e^{i\Delta\phi}\ket{\psi}$, along the meridian of the Bloch sphere corresponding to the fixed azimuthal angle $\phi$. The overall cyclic evolution introduces a parallel transport of the state $\ket{\psi}$ along an orange slice shaped path. The solid angle $\Delta\phi$ subtended by the orange slice shaped path, is the associated non-adiabatic Abelian geometric phase.}
\label{fig:SQG}
\end{figure}

We illustrate the above three steps evolution on the Bloch sphere in Fig. \ref{fig:SQG}. As shown in Fig. \ref{fig:SQG}, the first and third steps evolve the qubit state along the meridian of the Bloch sphere corresponding to the fixed azimuthal angle $\phi$. It is important to note that the parameters $\theta$ and $\phi$ are constant during the evolution and the magnetic strength $B$ is the only allowed time-dependent control variable. At the completion of the three steps, we have a cyclic evolution 
\begin{eqnarray}
\mathcal{U}(0, \tau_{3})\ket{\psi}=\mathcal{U}(\tau_{2}, \tau_{3})\mathcal{U}(\tau_{1}, \tau_{2})\mathcal{U}(0, \tau_{1})\ket{\psi}=e^{i\Delta\phi}\ket{\psi}.\nonumber\\
\label{GP}
\end{eqnarray} 
In fact,  this three-step evolution introduces a cyclic evolution of the general qubit state $\ket{\psi}$ about an orange slice shaped path on the Bloch sphere, where the two geodesic edges of the path differ as $\Delta\phi=\tilde{\phi}-\phi$ in their azimuthal angles.

Note that the evolution in step 1 satisfies the condition $(i)$ of Eq. (\ref{SDC}) and the evolutions in step 2 and 3 satisfy the condition $(ii)$ in Eq. (\ref{SDC}), which indicate that no dynamical phases occur along the three step evolutions. Therefore, the dynamical phase vanishs in the cyclic evolution of the general state $\ket{\psi}$ and the overall phase accumulated in this evolution, i.e., $\Delta\phi$ , is all geometric. Strictly speaking, the phase $\Delta\phi$, which is the solid angle subtended by the orange slice shaped path, is the non-adiabatic Abelian geometric phase accompanying the parallel transport of the state $\ket{\psi}$ along this orange slice shaped path \cite{aharonov1987}.

One may further note that the orthogonal counterpart state of $\ket{\psi}$, i.e., 
\begin{eqnarray}
\ket{\psi^{\bot}}=\sin\frac{\theta}{2}\ket{0}-e^{i\phi}\cos\frac{\theta}{2}\ket{1},
\end{eqnarray}
 would accordingly evolve in a cyclic fashion giving rise to 
\begin{eqnarray}
\mathcal{U}(0, \tau_{3})\ket{\psi^{\bot}}=e^{-i\Delta\phi}\ket{\psi^{\bot}}.
\label{OGP}
\end{eqnarray} 
with geometric phase $-\Delta\phi$.

Eqs. (\ref{GP}) and (\ref{OGP}) imply that the final time evolution operator $\mathcal{U}(0, \tau_{3})$ has actually a geometric structure given by 
\begin{eqnarray}
\mathcal{U}(0, \tau_{3})&=&\mathcal{U}(\tau_{2}, \tau_{3})\mathcal{U}(\tau_{1}, \tau_{2})\mathcal{U}(0, \tau_{1})\nonumber\\
&=&e^{i\Delta\phi}\ket{\psi}\bra{\psi}+e^{-i\Delta\phi}\ket{\psi^{\bot}}\bra{\psi^{\bot}},\nonumber\\
\end{eqnarray} 
which takes the following form in the qubit computational $\{\ket{0}, \ket{1}\}$ basis 
\begin{eqnarray}
\mathcal{U}(0, \tau_{3})=\mathcal{R}_{\vec{m}}(\Delta\phi)=\cos\Delta\phi+i\sin\Delta\phi[\vec{m}\cdot\vec{\sigma}]
\label{usqg}
\end{eqnarray} 
with $\vec{m}=(\sin\theta\cos\phi, \sin\theta\sin\phi, \cos\theta)$.  The Eq. (\ref{usqg}) indicates that the time evolution operator $\mathcal{U}(0, \tau_{3})$ is actually a general $SU(2)$ rotation about the rotation axis $\vec{m}$ with rotation angle given by non-adiabatic Abelian geometric phase $\Delta\phi$. Thus, the proposed evolution $\mathcal{U}(0, \tau_{3})$ introduces a practical route to realize universal non-adiabatic geometric single-qubit gates.

\subsection{Two-Qubit Gates}
A two-qubit gate on given register qubits $k$ and $l$ in the system is achieved by the two-qubit Hamiltonian $H_{kl}$ identified in Eq. (\ref{tqh}). Hamiltonian $H_{kl}$ embeds the computation system of two register qubits $k$ and $l$ into a host three-qubit system via, as shown in Fig. \ref{fig:star-shape}, indirect coupling of our register qubits $k$ and $l$ through a third auxiliary qubit $a$. The $H_{kl}$ in the computational basis takes the following double $\Lambda$ structure 
\begin{eqnarray}
H_{kl}&=&\frac{J_{l}}{2}\ket{010}\bra{001}+\frac{J_{k}}{2}\ket{010}\bra{100}+\nonumber\\
&&\frac{J_{k}}{2}\ket{101}\bra{011}+\frac{J_{l}}{2}\ket{101}\bra{110}+h.c.,
\label{lvtqh}
\end{eqnarray}
where $0$ and $1$ at each site in the basis states from left to right, respectively, represent the states of qubits $k$, $a$, $l$.      

Assume 
\begin{eqnarray}
(J_{k}, J_{l})=\Omega(\cos\frac{\theta}{2}, \sin\frac{\theta}{2}),
\end{eqnarray}
where $\Omega=\sqrt{J_{k}^{2}+J_{l}^{2}}$. If we fix the angle $\theta$ and turn on the Hamiltonian $H_{kl}$ for a time interval $[0, \tau]$ such that 
$\frac{1}{2}\int_{0}^{\tau}\Omega dt=\pi$ then the double $\Lambda$ structure of  $H_{kl}$ leads to the final time evolution operator
\begin{eqnarray}
\mathcal{U}(0,\tau)=e^{-i\int_{0}^{\tau}H_{kl}dt}=\mathcal{U}_{0}(0,\tau)\oplus\mathcal{U}_{1}(0,\tau)
\label{tqto}
\end{eqnarray}
in the ordered basis $\{\ket{000}, \ket{001}, \ket{100}, \ket{101}, \ket{010}, \ket{011},$ $\ket{110}, \ket{111}\}$, where
\begin{eqnarray}
\mathcal{U}_{0}(0,\tau)&=&\left(
\begin{array}{cccc}
1  & 0 & 0 & 0  \\
0 & \cos\theta & -\sin\theta & 0  \\
0  & -\sin\theta & -\cos\theta & 0  \\
0 & 0 & 0 & -1   
\end{array}
\right)\nonumber\\
\mathcal{U}_{1}(0,\tau)&=&\left(
\begin{array}{cccc}
-1  & 0 & 0 & 0  \\
0 & -\cos\theta & -\sin\theta & 0  \\
0  & -\sin\theta & \cos\theta & 0  \\
0 & 0 & 0 & 1   
\end{array}
\right).
\label{DSC}
\end{eqnarray}
We pursue with some remarks and properties of the system in Eq. (\ref{lvtqh}) and its time evolution operator described in Eq. (\ref{tqto}):
\begin{itemize}
\item Let us denote 
\begin{eqnarray}
\mathcal{H}_{q}=\text{span}\{\ket{0q0}, \ket{0q1}, \ket{1q0}, \ket{1q1}\}
\label{tqhsq}
\end{eqnarray}
$q=0, 1$. Each of the subspaces $\mathcal{H}_{q}$, $q=0, 1$, indeed corresponds to the four dimensional computational subspace of the two register qubits $k$ and $l$, when the auxiliary qubit is fixed to the state $\ket{q}$. Eq. (\ref{tqto}) indicates that the subspaces $\mathcal{H}_{q}$, $q=0, 1$, evolve in cyclic manners during the time interval $[0, \tau]$. Therefore the time evolution operator components $\mathcal{U}_{q}(0,\tau)$, $q=0,1$, introduce two-qubit gates on register qubits $k$ and $l$, if the auxiliary qubit is initialized and measured in the same basis state $\ket{q}$.

\item By evaluating the entangling powers \cite{zanardi00-r, balakrishnan10}
 \begin{eqnarray}
e_{p}[\mathcal{U}_{0}(0,\tau)]=e_{p}[\mathcal{U}_{1}(0,\tau)]=\frac{2}{9}[1-\cos^{4}\theta],
\end{eqnarray}
we obtain that both gates provide the same and full entangling power controlled by the parameter $\theta$. For each $\theta$ satisfying $|\cos\theta|<1$, the gates 
$\mathcal{U}_{q}(0,\tau)$, $q=0, 1$, are entangling two-qubit gates and thus they allow for universal quantum information processing when accompanied with universal single-qubit gates given in Eq. (\ref{usqg}).

\item Moreover, we notice that the gates $\mathcal{U}_{q}(0, \tau)$, $q=0,1$, have remarkable holonomic natures, which can be verified from two points of views. 

{\it First}: As mentioned in the first remark above, from Eq. (\ref{tqto}) we have that the subspaces $\mathcal{H}_{q}$ evolve in cyclic fashions during the time interval $[0, \tau]$. These cyclic evolutions actually take place in the Grassmanian $G(8,4)$, the space of all four dimensional subspaces of the eight dimensional Hilbert space of the three qubits $k$, $l$, and $a$. We may call $\mathcal{C}_{q}$ the corresponding loops in the Grassmanian $G(8,4)$. In addition, for each $q=0, 1$, one may observe that 
 \begin{eqnarray}
\mathcal{U}(0, t)\mathbf{P}_{q}\mathcal{U}^{\dagger}(0, t)H_{kl}\mathcal{U}(0, t)\mathbf{P}_{q}\mathcal{U}^{\dagger}(0, t)=0,
\label{zdp}
\end{eqnarray}
at each time $t\in [0, \tau]$, where $\mathbf{P}_{q}$ is the projection operator on the subspace $\mathcal{H}_{q}$ and $\mathcal{U}(0, t)=\exp[-i\int_{0}^{t}H_{kl}ds]$ is the evolution operator at time $t$. The Eq. (\ref{zdp}) follows from $[H_{kl}, \mathcal{U}(0, t)]=0$ at each time $t$ and that $\mathbf{P}_{q}H_{kl}\mathbf{P}_{q}=0$.

Therefore, the Eqs. (\ref{tqto}) and (\ref{zdp}) imply that for each $q=0,1$, the subspace $\mathcal{H}_{q}$ evolves cyclicly about the corresponding closed path $\mathcal{C}_{q}$ in the Grassmanian $G(8,4)$, along which no dynamical phases occur \cite{anandan88}. Mathematically speaking, the gate operator $\mathcal{U}_{q}(0, \tau)$, which is actually the projection of the final time evolution operator $\mathcal{U}(0, \tau)$ into the subspace $\mathcal{H}_{q}$, i.e., $\mathcal{U}_{q}(0, \tau)=\mathbf{P}_{q}\mathcal{U}(0, \tau)\mathbf{P}_{q}$, is the non-Abelian nonadiabatic quantum holonomy associated with the parallel transport of $\mathcal{H}_{q}$ about the loop $\mathcal{C}_{q}$ in the Grassmanian $G(8,4)$ \cite{anandan88}.

{\it Second}: Looking more carefully into the final time evolution operator given by Eqs. (\ref{tqto}, \ref{DSC}) and the double $\Lambda$ coupling structure of Eq. (\ref{lvtqh}),
we see that the two-qubit entangling gates $\mathcal{U}_{q}(0, \tau)$ possess even richer holonomic structures. For the sake of simplicity, in the following we restrict ourselves to
explain the further holonomic structure of the gate $\mathcal{U}_{0}(0, \tau)$, however the same type of explanation would exists for the gate $\mathcal{U}_{1}(0, \tau)$. 

We shall rewrite the two-qubit computational space $\mathcal{H}_{0}$ given in Eq. (\ref{tqhsq}) in the following directsum form 
 \begin{eqnarray}
\mathcal{H}_{0}=\mathcal{H}_{0}^{0}\oplus\mathcal{H}_{0}^{2}\oplus\mathcal{H}_{0}^{1},
\label{dsd}
\end{eqnarray}
where
$\mathcal{H}_{0}^{0}=\text{span}\{\ket{000}\}$,
$\mathcal{H}_{0}^{2}=\text{span}\{\ket{001}, \ket{100}\}$ and
$\mathcal{H}_{0}^{1}=\text{span}\{\ket{101}\}$. Accordingly, we may put the gate operator $\mathcal{U}_{0}(0, \tau)$ in a directsum form as 
\begin{eqnarray}
\mathcal{U}_{0}(0, \tau)=(1)\oplus\left(
\begin{array}{cc}
\cos\theta  & -\sin\theta  \\
-\sin\theta & -\cos\theta    
\end{array}
\right).\oplus(-1).
\label{dsdg}
\end{eqnarray}

The state $\ket{000}$ does not contribute into the Hamiltonian given in Eq. (\ref{lvtqh}) and thus it is kept unchanged during the time evolution of the system. In other words,
the evolution of the subspace $\mathcal{H}_{0}^{0}$ would be stationary with associated trivial phase during any time interval. This explains the first element, $(1)$, in the right hand side directsum of Eq. (\ref{dsdg}). 

However, the double $\Lambda$ structure of Eq. (\ref{lvtqh}) implies that the evolutions of $\mathcal{H}_{0}^{1}$ and $\mathcal{H}_{0}^{2}$ are non-stationary and, respectively, take place in the three dimensional invariant subspaces $\text{span}\{\ket{101}, \ket{011}, \ket{110}\}$ and $\text{span}\{\ket{001}, \ket{010}, \ket{100}\}$. Explicitly speaking, for each $d=1, 2$, the evolution of $\mathcal{H}_{0}^{d}$ specifies a non-trivial path in the Grassmanian $G(3,d)$, which we may here call $\mathcal{C}_{0}^{d}$. Moreover,  the block diagonal form of the final time evolution operator, $\mathcal{U}(0, \tau)$, in the corresponding ordered basis given below Eq. (\ref{tqto}) further implies that each of the subspaces $\mathcal{H}_{0}^{d}$, $d=1, 2$, undergoes a cyclic evolution during the time interval $[0, \tau]$ and thus the corresponding path $\mathcal{C}_{0}^{d}$ is a closed path in $G(3,d)$.
If we assume $\mathbf{P}_{0}^{d}$ to be the projection operator on the subspace $\mathcal{H}_{0}^{d}$ then from $\mathbf{P}_{0}^{d}H_{kl}\mathbf{P}_{0}^{d}=0$ and $[H_{kl}, \mathcal{U}(0, t)]=0$ we obtain 
\begin{eqnarray}
\mathcal{U}(0, t)\mathbf{P}_{0}^{d}\mathcal{U}^{\dagger}(0, t)H_{kl}\mathcal{U}(0, t)\mathbf{P}_{0}^{d}\mathcal{U}^{\dagger}(0, t)=0 
\end{eqnarray}
at each time $t$, for $d=1,2$, which indicates no dynamical phases occur along the cyclic evolutions $\mathcal{C}_{0}^{d}$, $d=1, 2$ \cite{anandan88}. All these verify that the subspaces $\mathcal{H}_{0}^{d}$, $d=1,2$, are actually parallel transported about the loops $\mathcal{C}_{0}^{d}$ giving rise to the following nonadiabatic quantum holonomies \cite{anandan88}

\begin{eqnarray}
U(\mathcal{C}_{0}^{1})&=&\mathbf{P}_{0}^{1}\mathcal{U}(0, t)\mathbf{P}_{0}^{1}=(-1)\nonumber\\
U(\mathcal{C}_{0}^{2})&=&\mathbf{P}_{0}^{2}\mathcal{U}(0, t)\mathbf{P}_{0}^{2}=\left(
\begin{array}{cc}
\cos\theta  & -\sin\theta  \\
-\sin\theta & -\cos\theta    
\end{array}
\right).
\label{sh}
\end{eqnarray}
As a result, we see in Eq. (\ref{dsdg}) that the two-qubit entangling gate $\mathcal{U}_{0}(0, \tau)$ is indeed composed of the holonomies in Eq. (\ref{sh}), namely  
 \begin{eqnarray}
\mathcal{U}_{0}(0, \tau)=(1)\oplus U(\mathcal{C}_{0}^{2})\oplus U(\mathcal{C}_{0}^{1}).
\label{dtqg}
\end{eqnarray}

In conclusion, our analysis above reveals the rich holonomic nature of the two-qubit entangling gate $\mathcal{U}_{0}(0, \tau)$ by demonstrating that the gate $\mathcal{U}_{0}(0, \tau)$ not only as a whole is a nonadiabatic holonomy but also each of its nonzero block constitutes is a nonadiabatic holonomy.
\end{itemize}
\section{Discussion and summary}
As the nonadiabatic holonomies became an important approach for implementation of fast fault-tolerant quantum gates, experimental implementation of nonadiabatic holonomic quantum computation with spin qubits, as a natural and suitable platform for realization of quantum computers, caught increasing interests in recent years \cite{Arroyo-Camejo2014, Zu2014, Zhou2017, sekiguchi2017, ishida2018}.   Despite a number of significant efforts in this area, only single-qubit gates have been addressed. Therefore, still a lack of feasible scheme, which supports scalability as well as universal single-qubit and two-qubit entangling gates all in the same configuration is felt. Compared to the existing works, our scheme above is consist of arbitrary $n$ register spin qubits arranged in a star-shape architecture about a shared single auxiliary spin qubit in the middle (see Fig. \ref{fig:star-shape}). In addition to the scalability, the proposed star-shape configuration permits for universal nonadiabatic holonomic quantum computation, where an arbitrary holonomic single-qubit gate on each register qubit is achieved by a local transverse magnetic field and a two-qubit entangling gate between a given pair of register qubits is performed in a double $\Lambda$ structure as demonstrated by indirect bridge coupling between the selected register qubits through the auxiliary qubit. While single-qubit gates are realized through Abelian nonadiabatic holonomies \cite{aharonov1987}, the proposed entangling two-qubit gates obey a rich holonomic description associated with non-Abelian as well as Abelian nonadiabatic holonomies \cite{anandan88}. All holonomic universal computations take place in the subspace of the system, where the state of auxiliary qubit is fixed to one of its computational basis states (say for instance the basis state $\ket{0}$). A universal circuit corresponding to our nonadiabatic holonomic scheme is depicted in Fig. \ref{fig:uc}. 
\begin{widetext}
\begin{center}
\begin{figure}[h]
\begin{center}
\includegraphics[width=15cm]{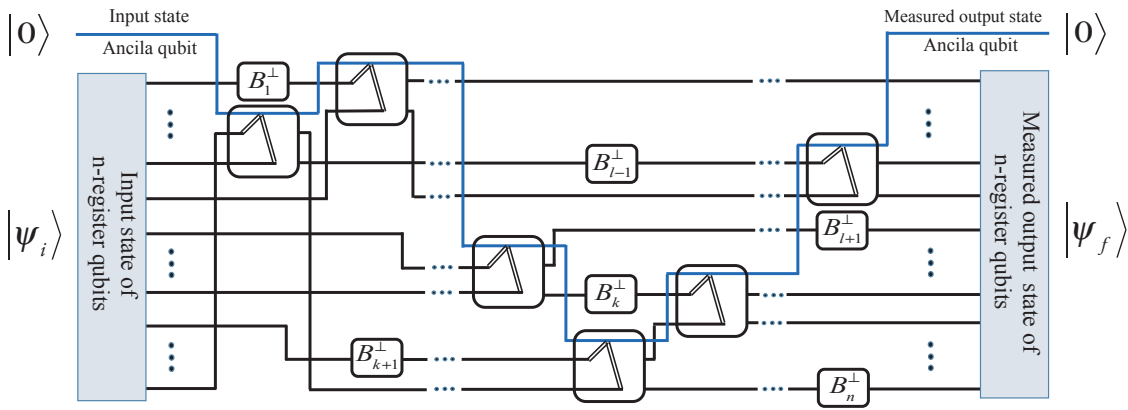}
\end{center}
\caption{(Color online) Schematic digram of a universal holonomic quantum computation. Holonomic single-qubit gates are implemented by local transverse magnetic fields, which introduce transverse coupling between two computational basis states of qubits. The auxiliary qubit, which is illustrated in blue, only contributes in two-qubit gates. Two register qubits are coupled through the auxiliary qubit in a double $\Lambda$ structure allowing for implementation of two-qubit entangling gates. The auxiliary qubit is initialized, and measured at the end of computation in the same computational basis state, which here we selected to be the state $\ket{0}$. }
\label{fig:uc}
\end{figure}
\end{center}
\end{widetext}

In summary, we have proposed a scalable spin-based setup for universal nonadiabatic holonomic quantum computation. We hope the present scheme helps to overcome practical challenges and establish a feasible platform for realization of scalable universal nonadiabatic holonomic quantum computation particularly with spin qubits.
The discussion for the holonomic nature of the gates would further improve our understanding of the concept of quantum holonomy in solid state systems and its relation to quantum computation.   

\section{Acknowledgment }
This work was supported by Department of Mathematics at University of Isfahan (Iran). 
The author acknowledges financial support from the Iran National Science Foundation (INSF) 
through Grant No. 96008297.
\bibliography{SSANHS}

\end{document}